# CODING BLOCK-LEVEL PERCEPTUAL VIDEO CODING FOR 4:4:4 DATA IN HEVC

*Lee Prangnell, Miguel Hernández-Cabronero and Victor Sanchez*

University of Warwick, Coventry, England, UK

**ABSTRACT**

There is an increasing consumer demand for high bit-depth 4:4:4 HD video data playback due to its superior perceptual visual quality compared with standard 8-bit subsampled 4:2:0 video data. Due to vast file sizes and associated bitrates, it is desirable to compress raw high bit-depth 4:4:4 HD video sequences as much as possible without incurring a discernible decrease in visual quality. In this paper, we propose a Coding Block (CB)-level perceptual video coding technique for HEVC named Full Color Perceptual Quantization (FCPQ). FCPQ is designed to adjust the Quantization Parameter (QP) at the CB level — i.e., the luma CB and the chroma Cb and Cr CBs — according to the variances of pixel data in each CB. FCPQ is based on the default perceptual quantization method in HEVC called AdaptiveQP. AdaptiveQP adjusts the QP of an entire $2N{\times}2N$ CU based only on the spatial activity of the constituent luma CB. As demonstrated in this paper, by not accounting for the spatial activity of the constituent chroma CBs, as is the case with AdaptiveQP, coding performance can be significantly affected; this is because the variance of pixel data in a luma CB is notably different from the variances of pixel data in chroma Cb and Cr CBs. FCPQ, therefore, addresses this problem. In terms of coding performance, FCPQ achieves BD-Rate improvements of up to 39.5% (Y), 16% (Cb) and 29.9% (Cr) compared with AdaptiveQP.

*Index Terms* — Perceptual Video Coding, HEVC, 4:4:4, Perceptual Quantization, Coding Block

## 1. INTRODUCTION

Profound technological advancements have recently emerged in visual display hardware and video coding software. Consequently, this has given rise to an increasing consumer demand in relation to the playback of high quality video signals including high bit-depth 4:4:4 HD video data. Top-of-the-range TVs and Visual Display Units (VDUs) ubiquitously support the playback of compressed video data which possess the following technical characteristics: High Definition (HD), Ultra HD (UHD), High Dynamic Range (HDR), Wide Color Gamut (WCG), YCbCr/RGB 4:4:4 and high bit-depth video data. Furthermore, high-end contemporary 4K and 8K UHD TVs include a HEVC codec which enables the decoding and playback of HEVC compressed video signals.

A major issue with high bit-depth and full color (RGB and YCbCr 4:4:4) raw video data is the associated uncompressed file size. An example of this is the BirdsInCage 10-bit 4:4:4 RGB 1080p HD video sequence provided by JCT-VC. For 10 seconds of footage at 60 frames per second, the raw file size of this sequence is 7 gigabytes. The main reasons for this are as follows: 1) The sequence is not spatially subsampled, as is the case with chroma subsampled 4:2:0 raw video data; 2) Compared with 8-bit raw video data, 10-bit raw video data requires an extra two bits of data storage per pixel in each color channel. This inevitably results in vast file sizes for the uncompressed, raw video data.

In video coding and image coding standards, quantization is a key element in terms of compressing raw data. In the HEVC standard, the default quantization technique employed is Uniform Reconstruction Quantization (URQ) [1, 2]. URQ is utilized to quantize the residue generated after intra and/or inter prediction. With URQ, a quantization step size (QStep) value — to which the QP value is mapped — is applied to all DCT/DST transformed residual values (transform coefficients) in a Transform Block (TB). URQ is not a perceptually optimized quantization method because it is designed to equally quantize entire TBs of luma and chroma transform coefficients based on the QStep value.

In terms of perceptual quantization and the Human Visual System (HVS), numerous psychophysical experiments reveal that the HVS is less sensitive to quantization-related distortions within regions of image data that comprise high spatial variations [3]-[6]; this constitutes a form of visual masking in the spatial domain. In the context of video coding, this means that higher levels of quantization can be applied to high spatial activity regions in the frames of a sequence. Therefore, in comparison with standard uniform quantizers including URQ, perceptual quantizers can be designed to exploit the spatial domain visual masking phenomenon of the HVS, thus potentially giving rise to bitrate reductions. For example, this can be achieved by applying coarser levels of quantization to regions in video data that comprise high spatial activity (as quantified by the variance of pixel values).

In HEVC, AdaptiveQP [7]-[10] is an example of a HVS-based perceptual quantization method that can exploit the spatial domain visual masking phenomenon of the HVS. A notable shortcoming of AdaptiveQP, however, is that it adjusts the QP of an entire $2N{\times}2N$ CU based solely on the pixel variance in the constituent luma CB [7]-[10]; FCPQ is designed to address this problem. If AdaptiveQP is utilized to decrease overall bitrates by increasing the CU-level QP (based only on the pixel variance in a luma CB), an inappropriate QP adjustment may be applied to the chroma Cb and Cr CBs; this is particularly pertinent to high bit-depth 4:4:4 video data. The chroma Cb and Cr channels in, for example, 10-bit 4:4:4 video sequences may contain higher pixel variances compared with the pixel variances in the corresponding luma channel. This is by virtue of the nature of 10-bit 4:4:4 video data; i.e., 30 bits per pixel (up to $1024^3$ unique colors per pixel) in addition to an absence of chroma subsampling. Therefore, for the variance-based perceptual quantization of high bit-depth 4:4:4 video data, separately adjusting the QP for the Y CB, the Cb CB and the Cr CB is significantly desirable.

Other luma-based, spatial domain perceptual quantization methods, similar to AdaptiveQP, have been previously proposed. The IDSQ technique in [11] is based on luminance masking; it is modeled on a Just Noticeable Distortion (JND) approach. The technique in [12] is similar to IDSQ; it is designed for the perceptual quantization of HDR video data. To the best of our knowledge, full color, CB-level perceptual quantization has not been previously explored in HEVC research.

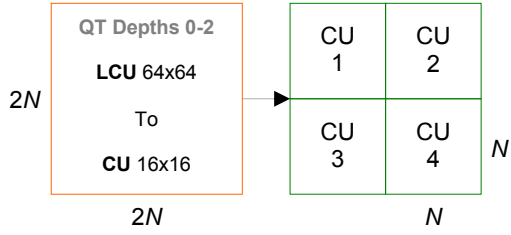

**Fig. 1:** The $2N \times 2N$ CUs at QT Depth Levels 0-2 are partitioned into four constituent $N \times N$ CUs; $N=32$ (level 0), $N=16$ (level 1) and $N=8$ (level 2). FCPQ and AdaptiveQP operate at these levels.

In this paper, we propose FCPQ, a full color perceptual quantization contribution; in contrast to AdaptiveQP, FCPQ is designed to perceptually adjust the QP at the CB level. FCPQ accounts for the spatial activity (i.e., the pixel variances) of the chroma Cb and Cr CBs, in addition to the pixel variance of the luma CB, when making QP selections in a $2N \times 2N$ CU. In other words, FCPQ separately adjusts the QP for the Y CB, the Cb CB and the Cr CB based on the pixel variances in each CB. This has the potential to significantly improve coding performance in comparison with AdaptiveQP, particularly for high bit-depth 4:4:4 video data. This is because Y, Cb and Cr pixel variances in high bit-depth 4:4:4 video data are typically considerably different from the corresponding variances in 8-bit subsampled 4:2:0 video data. Compared with 8-bit subsampled 4:2:0 video data, 10-bit 4:4:4 video data contains a much greater number of color variations in each pixel (in all color channels). Therefore, the higher the bit depth of the video data, in addition to an absence of chroma subsampling, the more potential there is for the chroma Cb and Cr channels to comprise higher pixel variances; FCPQ exploits this.

The rest of this paper is organized as follows. Section 2 includes technical information on the AdaptiveQP method. Section 3 provides comprehensive technical details of the proposed FCPQ technique. Section 4 includes the evaluation, results and discussion of FCPQ. Finally, Section 5 concludes the paper.

## 2. ADAPTIVE-QP IN HEVC

In HEVC, the CU contains three CBs: one Y CB, one Cb CB and one Cr CB for video data that is not monochrome. Assuming that the split flag is enabled, the $2N \times 2N$ CU comprises four constituent $N \times N$ CUs (see Fig. 1). The Largest Coding Unit (LCU) supports up to $64 \times 64$ samples and the Smallest Coding Unit (SCU) supports up to $8 \times 8$ samples. Note that LCUs operate at QuadTree (QT) Depth Level 0 and SCUs operate at QT Depth Level 3 [13]-[15].

AdaptiveQP is a $2N \times 2N$ CU-level perceptual quantization technique [7]-[10]. It can exploit the high spatial activity visual masking phenomenon of the HVS by employing a higher QP — relative to the URQ QP — to an entire CU if the CU contains a high spatial activity luma CB in which the pixel variance is high. This CU-level higher QP selection can improve coding performance compared with URQ [7, 10]. AdaptiveQP can also decrease the CU-level QP — relative to the URQ QP — if low spatial activity regions are detected. This functionality is in place because the HVS is more sensitive to quantization-induced compression artifacts in smooth regions of heavily quantized video data. The CU-level lower QP selection in AdaptiveQP can yield improved reconstruction quality compared with URQ [7, 10]. To summarize, AdaptiveQP increases or decreases the QP of an entire $2N \times 2N$ CU based on the spatial activity of pixel data in the corresponding luma CB, but without taking into account the spatial activity of the pixel data in the chroma Cb and Cr CBs.

In the AdaptiveQP technique, the CU-level perceptual QP, denoted as $PQ_Y$, is computed in (1):

$$PQ_Y = Q + \left[ 6 \times \log_2(L) \right] \quad (1)$$

where $Q$ denotes the QP value (before perceptual adjustment) and where $L$ refers to the normalized spatial activity of a luma CB. $Q$ and $L$ are computed in (2)-(4), respectively:

$$Q = \left[ 6 \times \log_2(QStep) \right] + 4 \quad (2)$$

$$QStep = 2^{\frac{Q-4}{6}} \quad (3)$$

$$L = \frac{f \cdot l + t_Y}{l + f \cdot t_Y} \quad (4)$$

where $QStep$ denotes the quantization step size value. Variable $f$ refers to a scaling factor for normalizing the spatial activity of a luma CB [7], $l$ corresponds to the non-normalized spatial activity of a luma CB and $t_Y$ denotes the mean spatial activity of all $2N \times 2N$ luma CBs belonging to the current picture [7]. Variables $f$, $l$ and $t_Y$ are computed in (5)-(7), respectively:

$$f = 2^{\frac{A}{6}} \quad (5)$$

$$l = 1 + \min\left(\sigma^2_{Y,d}\right), \quad \text{where } d = 1,\ldots,4 \quad (6)$$

$$t_Y = \frac{1}{C_Y} \sum_{n=1}^{C_Y} l_n \quad (7)$$

where $\sigma^2_{Y,d}$ denotes the variance of pixel values in the constituent $N \times N$ sub-block $d$ of a luma CB. Variable $C_Y$ refers to the number of $2N \times 2N$ luma CBs in the current picture and $A$ denotes the maximum allowable difference of the CU-level adjusted QP from the URQ QP. Note that $A = 6$ is the default value configured in HEVC HM 16 [7, 15]. Variable $\sigma^2_{Y,d}$ is computed in (8):

$$\sigma^2_{Y,d} = \frac{1}{m_Y} \sum_{n=1}^{m_Y} \left( w_n - \mu_Y \right)^2 \quad (8)$$

where $m_Y$ corresponds to the number of pixel values in sub-block $d$, $w_n$ denotes the $n^{th}$ pixel in sub-block $d$ and $\mu_Y$ refers to the mean pixel value of sub-block $d$.

## 3. PROPOSED FCPQ TECHNIQUE FOR HEVC

FCPQ expands on our previously proposed method, known as Color-Based Adaptive Quantization (C-BAQ) [16], which is a cross-color channel CU-level perceptual quantization technique based on the technical principles of AdaptiveQP. C-BAQ adjusts the QP of an entire $2N \times 2N$ CU by computing the sum of all pixel variances in the corresponding luma and chroma CBs, which equates to cross-color channel dependency for QP selection; FCPQ is designed to improve upon C-BAQ. By separately computing the variances of the pixel data in the luma CB and the chroma Cb and Cr CBs, this equates to a more accurate and also a more refined approach in terms of adjusting the QP according to spatial activity. Furthermore, in the standardized Format Range Extensions (RExt) of HEVC, JCT-VC (ITU-T/ISO/IEC) has provided the flexibility for signaling chroma QP offsets at the CB level in the Picture Parameter Set (PPS) in HEVC HM [17, 18]. FCPQ takes advantage of this flexibility, whereas C-BAQ and AdaptiveQP do not.

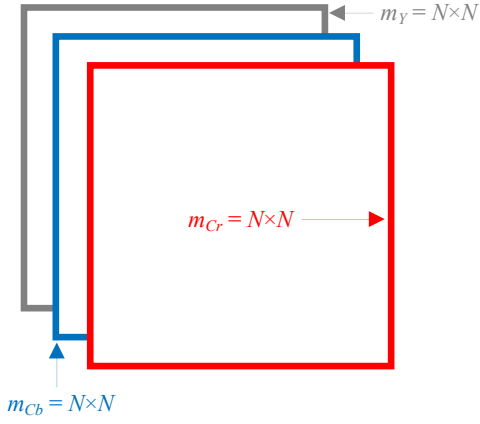

**Fig. 2:** The size of the constituent Y (Gray), Cb (Blue) and Cr (Red) CB sub-blocks, denoted as $d$, $k$ and $z$, respectively, in FCPQ for 4:4:4 video data.

FCPQ separately adjusts the QPs for the luma CB and the chroma Cb and Cr CBs based on the variances of the pixel data in the constituent sub-blocks of all three CBs (see Fig. 2). This functionality is useful for perceptually compressing high bit-depth 4:4:4 video data. Furthermore, due to the variance-based nature of FCPQ, the proposed method is, by definition, compatible with ITU-T BT.2020 standardized UHD and HDR/WCG content.

To reiterate, FCPQ is designed primarily to exploit the aforementioned visual masking phenomenon of the HVS — for high spatial activity regions in a picture — by increasing quantization levels for high variance luma and chroma CBs. The increased levels of quantization will result in bitrate reductions without incurring a noticeable decrease in perceptual quality. FCPQ can also decrease the QP if the luma and chroma CBs are calculated as having low variance values. Therefore, when a low spatial activity CB is detected, the CB-level QP is decreased, thus potentially leading to reconstruction quality improvements.

Recall that AdaptiveQP accounts for the spatial activity in a luma CB; FCPQ utilizes this functionality. Consequently, the operations described in equations (1)-(8) are employed in FCPQ. The chroma Cb and Cr CB-level perceptual QPs, denoted as $PQ_{Cb}$ and $PQ_{Cr}$, are computed in (9) and (10), respectively:

$$PQ_{Cb} = Q + \left\lceil 6 \times \log_2(B) \right\rceil \quad (9)$$

$$PQ_{Cr} = Q + \left\lceil 6 \times \log_2(R) \right\rceil \quad (10)$$

where $B$ and $R$ refer to the normalized spatial activity of chroma Cb and Cr CBs, respectively. $B$ and $R$ are computed in (11) and (12), respectively:

$$B = \frac{f \cdot b + t_{Cb}}{b + f \cdot t_{Cb}} \quad (11)$$

$$R = \frac{f \cdot r + t_{Cr}}{r + f \cdot t_{Cr}} \quad (12)$$

where $b$ and $r$ denote the non-normalized spatial activity of chroma Cb and Cr CBs, respectively. Variables $t_{Cb}$ and $t_{Cr}$ refer to the mean spatial activity of all $2N \times 2N$ chroma Cb and Cr CBs belonging to the current picture, respectively. Variables $b$, $r$, $t_{Cb}$ and $t_{Cr}$ are computed in (13)-(16), respectively:

$$b = 1 + \min\left(\sigma^2_{Cb,k}\right), \quad \text{where } k = 1,\ldots,4 \quad (13)$$

$$t_{Cb} = \frac{1}{C_{Cb}} \sum_{n=1}^{C_{Cb}} b_n \quad (14)$$

$$r = 1 + \min\left(\sigma^2_{Cr,z}\right), \quad \text{where } z = 1,\ldots,4 \quad (15)$$

$$t_{Cr} = \frac{1}{C_{Cr}} \sum_{n=1}^{C_{Cr}} r_n \quad (16)$$

where $\sigma^2_{Cb,k}$ and $\sigma^2_{Cr,z}$ refer to the variances of pixel values in the $N \times N$ chroma Cb sub-block $k$ and the chroma Cr sub-block $z$, respectively, of the chroma Cb and Cr CBs, respectively. Variables $C_{Cb}$ and $C_{Cr}$ denote the number of $2N \times 2N$ chroma Cb CBs and Cr CBs in the current picture, respectively. Variables $\sigma^2_{Cb,k}$ and $\sigma^2_{Cr,z}$ are quantified in (17) and (18), respectively:

$$\sigma^2_{Cb,k} = \frac{1}{m_{Cb}} \sum_{n=1}^{m_{Cb}} \left(v_n - \mu_{Cb}\right)^2 \quad (17)$$

$$\sigma^2_{Cr,z} = \frac{1}{m_{Cr}} \sum_{n=1}^{m_{Cr}} \left(j_n - \mu_{Cr}\right)^2 \quad (18)$$

where $m_{Cb}$ and $m_{Cr}$ denote the number of pixel values in sub-blocks $k$ and $z$, respectively (see Fig. 2), where $v_n$ and $j_n$ refer to the $n^{th}$ pixel values in sub-blocks $k$ and $z$, respectively, and where $\mu_{Cb}$ and $\mu_{Cr}$ correspond to the mean pixel values of sub-blocks $k$ and $z$, respectively.

## 4. EXPERIMENTAL EVALUATION AND DISCUSSION

Due to the fact that FCPQ is a HVS-based perceptual method, it is necessary to undertake subjective visual quality inspections of the reconstructed sequences. As such, we integrate FCPQ into HEVC HM 16.7 [15] and engage in several visual inspections in addition to performing a comprehensive objective evaluation. The primary goal of the informal subjective evaluation is to fairly assess the perceptual efficacy of FCPQ compared with AdaptiveQP. In terms of the objective evaluation, we follow, as closely as possible, the Common Test Conditions and Software Reference Configurations, as recommended by JCT-VC [20]. The visual inspection entails four participants engaging in a comprehensive analysis of the reconstruction quality of the FCPQ coded sequences, compared with the AdaptiveQP coded sequences, in order to establish if visual differences can be discerned.

Identical experimental conditions are applied to both FCPQ and AdaptiveQP in order to ensure fair testing. The proposed method is evaluated on 12 official sequences provided by JCT-VC (the YCbCr and RGB versions of six different sequences). The video sequences comprise the following technical characteristics: 10-bit 4:4:4 HD (1080p). In terms of the objective evaluation, and in accordance with the Common Test Conditions and Software Reference Configurations (as recommended by JCT-VC), we employ the standard four QP data points (i.e., initial QPs 22, 27, 32 and 37) in order to compute the Bjøntegaard Delta Rate (BD-Rate). BD-Rate is a metric that quantifies a technique's coding efficiency performance when the PSNR is computed as the same, on average, over four QP data points (in our case, initial QPs 22, 27, 32 and 37) [21]. The All Intra (AI) and Random Access (RA) encoding configurations are employed in the experimental setup, which includes the Main_444_10_Intra and Main_444_10 encoding profiles, respectively.

**Table 1:** YCbCr and RGB BD-Rate percentage reductions (i.e., performance improvements) attained by the proposed FCPQ technique compared with AdaptiveQP. The All Intra test results are shown on the left and the Random Access test results are shown on the right.

| FCPQ versus AdaptiveQP (YCbCr and RGB 4:4:4) – All Intra | | | | | | FCPQ versus AdaptiveQP (YCbCr and RGB 4:4:4) – Random Access | | | | | | |
|---|---|---|---|---|---|---|---|---|---|---|---|---|
| | YCbCr BD-Rate (%) | | | RGB BD-Rate (%) | | | | YCbCr BD-Rate (%) | | | RGB BD-Rate (%) | |
| | Y | Cb | Cr | G | B | R | | Y | Cb | Cr | G | B | R |
| BirdsInCage | −17.3 | −9.3 | −13.9 | −18.9 | −10.1 | −17.0 | BirdsInCage | −13.8 | −12.2 | −13.8 | −11.3 | −11.1 | −12.4 |
| DuckAndLegs | −14.7 | −5.3 | −10.2 | −6.2 | −4.9 | −6.8 | DuckAndLegs | −16.9 | −13.8 | −14.5 | −9.7 | −7.8 | −11.1 |
| Kimono | −20.8 | −4.6 | −20.7 | −19.3 | −7.8 | −14.8 | Kimono | −15.5 | −14.2 | −14.1 | −8.6 | −9.3 | −10.0 |
| OldTownCross | −33.2 | 2.6 | −12.3 | −18.3 | −6.7 | −11.4 | OldTownCross | −39.5 | −16.0 | −29.9 | −17.8 | −18.7 | −17.8 |
| ParkScene | −17.1 | −3.7 | −14.6 | −13.3 | −4.1 | −11.3 | ParkScene | −12.5 | −12.4 | −11.7 | −8.6 | −9.3 | −10.0 |
| Traffic | −12.7 | −6.3 | −8.9 | −10.6 | −3.8 | −7.0 | Traffic | −5.6 | −7.1 | −5.7 | −1.5 | −4.3 | −0.9 |

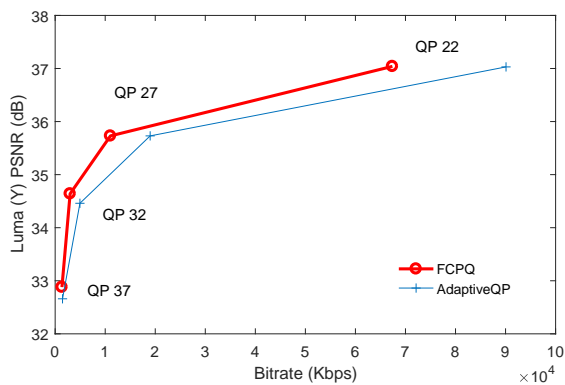

**Fig. 3:** FCPQ bitrate reductions (Y channel) compared with AdaptiveQP on the 10-bit 4:4:4 YCbCr sequence OldTownCross.

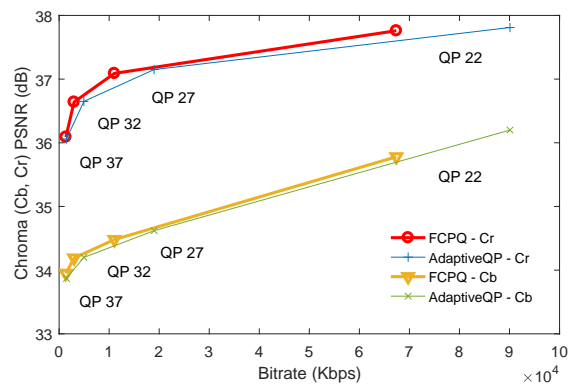

**Fig. 4:** FCPQ bitrate reductions (Cb, Cr channels) compared with AdaptiveQP on the 10-bit 4:4:4 YCbCr sequence OldTownCross.

In the subjective viewings, four participants carried out 48 visual inspections on the FCPQ coded sequences and the AdaptiveQP coded sequences. Initial QP 37 ($Q = 37$) is employed to ensure that any potential compression artifacts are more easily discernible. In line with ITU-T P.910 [22] subjective evaluation recommendations, spatial and temporal visual fidelity assessments were performed at various viewing distances. Furthermore, the coded sequences were viewed on HD 1080p and 2560×1440 (WQHD) resolution displays. Out of the 48 visual inspections, 90% reported no visual quality differences between the FCPQ coded sequences and the AdaptiveQP coded sequences (see Fig. 5).

In the tests conducted, FCPQ yields important luma and chroma BD-Rate improvements (see Table 1). The proposed technique is considerably effective on the YCbCr version of the OldTownCross sequence; FCPQ achieves BD-Rate reductions of 39.5% (Y), 16% (Cb) and 29.9% (Cr) in the RA tests (see Table 1, Fig. 3 and Fig. 4). Focusing on the OldTownCross RA tests, a high compression performance is attained because of the distribution of pixel data in the sequence. Consequently, high spatial variances are detected in the luma CBs and the chroma Cb and Cr CBs. FCPQ, therefore, increases the QPs, relative to the URQ QPs, at the luma and chroma CB level, thus giving rise to important BD-Rate reductions. Furthermore, certain regions within the OldTownCross sequence consist of low variance pixel data; therefore, low spatial variances are detected in the luma CBs and the chroma Cb and Cr CBs. Accordingly, FCPQ decreases the QPs, relative to the URQ QPs, at the luma and chroma CB level. For this reason, PSNR value increases are attained in some cases (see the plots in Fig. 3 and Fig. 4). In relation to computational complexity including encoding times and decoding times, no significant differences between FCPQ and AdaptiveQP are observed.

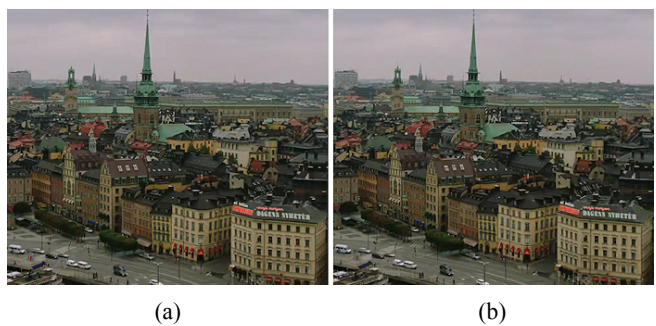

(a)                                    (b)

**Fig. 5:** The 10-bit 4:4:4 HD sequence OldTownCross (YCbCr). Subfigure (a): coded with the proposed FCPQ method ($Q = 37$ and RA — bitrate: 1355.65 Kbps for 150 frames). Subfigure (b): coded with AdaptiveQP ($Q = 37$ and RA — bitrate: 1472.44 Kbps for 150 frames).

## 5. CONCLUSION

FCPQ is proposed to potentially replace AdaptiveQP in HEVC for the perceptual compression of high bit-depth 4:4:4 video data. AdaptiveQP is a CU-level perceptual quantization method that only accounts for the pixel variance in a luma CB when applying a CU-level QP adjustment; FCPQ improves upon this by adjusting the QP at the CB level. In FCPQ, CB-level QP adjustments are achieved by employing a visual masking approach based on computing the variance of pixel data in the Y CB, the Cb CB and the Cr CB. Compared with AdaptiveQP, FCPQ achieves BD-Rate reductions of up to 39.5% (Y), 16% (Cb) and 29.9% (Cr) with no appreciable visual differences in the reconstructed sequences.